\documentclass[letter,10pt,twocolumn]{article}
\usepackage{float}
\usepackage{graphicx}
\usepackage{dcolumn}
\usepackage{amsmath}
\usepackage[dvips]{epsfig}
\usepackage{subfigure}

\begin{document}

\title{Design of photonic crystal microcavities for cavity QED}
\author{Jelena Vu\v{c}kovi\'{c}, Marko Lon\v{c}ar, Hideo Mabuchi, and Axel
Scherer\\California Institute of Technology, Mail Code 136-93,
Pasadena, CA 91125}
\date{\today}
\maketitle

\begin{abstract}
\vbox{\vspace{0.1in}} We discuss the optimization of optical
microcavity designs based on 2-d photonic crystals for the purpose
of strong coupling between the cavity field and a single neutral
atom trapped within a hole. We present numerical predictions for
the quality factors and mode volumes of localized defect modes as
a function of geometric parameters, and discuss some experimental
challenges related to the coupling of a defect cavity to gas-phase
atoms.
\end{abstract}

\section{Introduction}

A variety of passive and active optical devices can be constructed
by introducing point or line defects into a periodic array of
holes perforating an optically thin semiconductor slab. In such
structures, light is confined within the defect regions by the
combined action of distributed Bragg reflection and internal
reflection. This technique has been employed in making a
microcavity semiconductor (InGaAsP) laser \cite{ref:Painter99}
(emitting at $\lambda=1.55$ $\mu$m), and for demonstrating Si
optical waveguides with sharp bends \cite{ref:Loncar00}. One can
thus easily envision the fabrication of integrated optical
``networks'' on a single chip, with numerous microcavity-based
active devices linked by passive waveguide interconnects.

The combination of high quality factor and extremely low mode
volume that should be obtainable in point-defect microcavities
makes the photonic crystal (PC) paradigm especially attractive for
experiments in cavity quantum electrodynamics (cavity QED)
\cite{ref:Kimb94}, with potential applications in quantum
information technology. The primary focus of this paper will be to
discuss the optimization of microcavity designs for cavity QED
with strong coupling between defect modes and gas-phase neutral
atoms. This new paradigm for cavity QED poses formidable technical
challenges with regard to atom trapping and the characterization
of surface effects, so we include some discussion of these issues
and of our current approaches to addressing them.

In Section II of this article we begin by analyzing elementary
microcavities formed by changing either the refractive index or
radius of a single defect hole. We discuss in detail some problems
in previous FDTD calculations done by our group
\cite{ref:Painter98}, as intuitions gained from the resolution of
these problems are important in the microcavity design
optimization for cavity QED. In Section III we present our designs
of microcavities optimized for strong coupling between the cavity
field and an atom trapped within a hole of the PC. Based on
numerical analysis using the 3D finite-difference time-domain
(FDTD) method, we predict that quality factors over $3\times 10^4$
can be achieved in these structures. We also discuss in this
section the difficulty of estimating surface interactions between
a trapped atom and the semiconductor substrate, and propose a
technical strategy for confining gas-phase atoms within a defect
microcavity. Finally, in Section IV we describe a procedure that
we have developed for microcavity fabrication.

\begin{figure}[htbp]
\begin{center}
\epsfig{figure=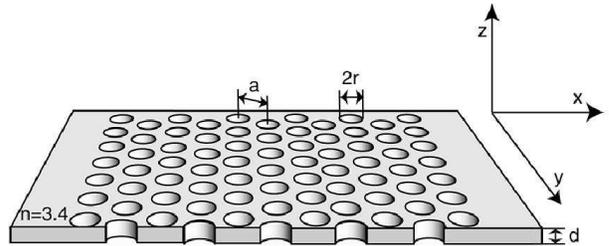, width=3.5in}
\caption{Optically thin slab patterned with a hexagonal array of air holes.} \label{fig:strucparams}
\end{center}
\end{figure}

\section{Single defect hole}

The microcavities analyzed in this paper are formed by introducing point defects into a dielectric slab patterned with a hexagonal array of air holes. The
thickness of the slab is $d$ and its refractive index is equal to $3.4$. The spacing between holes is denoted by $a$ and the hole radius by $r$, as shown in
Figure \ref{fig:strucparams}. We use $\lambda$ to specify the optical wavelength in air. In our calculations of microcavity quality factors, the boundary for
separation of vertical from lateral loss (i.e. the vertical quality factor $Q_{\perp}$ from the lateral quality factor $Q_{||}$) is positioned approximately at
$\lambda/2$ from the surface of the membrane. As the number of PC layers around a defect is increased, $Q_{||}$ increases and the total quality factor $Q$
approaches $Q_{\perp}$. We will adopt a coordinate convention in which $x=0$, $y=0$, $z=0$ denotes the center of the cavity and $z=0$ is the middle plane of the
slab.

The simplest method of forming a microcavity within the structure
shown in Figure \ref{fig:strucparams} is to change the radius or
index of refraction of a single hole. The former method is more
interesting from the perspective of fabrication, since the
lithographic tuning of geometric parameters of individual holes is
a simple process, but in this section we will consider both
methods. By increasing the radius of a single hole an acceptor
defect state is excited, {\it i.e.}, pulled into the band-gap from
the dielectric band. On the other hand, by decreasing the radius
of an individual hole (or by tuning its refractive index between 1
and the refractive index of the slab) a donor defect state is
pulled into the band-gap from the air band \cite{ref:Yablo91}.
Acceptors tend to concentrate their electric field energy density
in regions where the larger (semiconductor) refractive index was
located in the unperturbed PC,
while the electric field energy density of donors is concentrated
in regions where there was air in the unperturbed PC.
Since the electric field intensity in air regions is small in case
of acceptor defects, these are not good candidates for strong
coupling with a single gas-phase atom that would be trapped within
a hole. In this article we will thus focus on donor states. For a
discussion of acceptor states excited in an optically thin slab
perforated with a hexagonal PC array, readers are referred to
Ref.~\cite{ref:JV_spie2001}.

\begin{figure}[htbp]
\begin{center}
\epsfig{figure=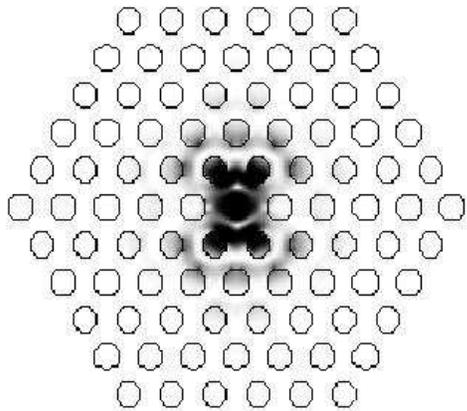, width=2.5in}
\caption{Electric field intensity pattern of the
$x$-dipole mode. Parameters of the structure are $r/a=0.3$, $d/a=0.6$ and in the used discretization $a=15$. The refractive index of the slab is $3.4$ and of the
defect is $2.4$. The plotted intensity patterns are for the $x-y$ plane at the middle of the slab.} \label{fig:xydn24}
\end{center}
\end{figure}

\subsection{Changing the refractive index of a single hole}

Microcavity formation by alteration of the refractive index of a single defect hole in a hexagonal PC has been analyzed previously by our group
\cite{ref:Painter98}. In that analysis, we predicted that dipole-like donor states (such as the $x$-dipole mode shown in Figure \ref{fig:xydn24}) with quality
factors up to $3\times 10^4$ should exist. We now believe that the quality factors of such microcavities are limited to several thousand, for reasons discussed
below. This discussion reveals the extreme sensitivity of microcavity quality factors to small distortions to the local PC geometry, which will later be used as
a powerful design tool in optimization for cavity QED.

\begin{figure}[htbp]
\begin{center}
\epsfig{figure=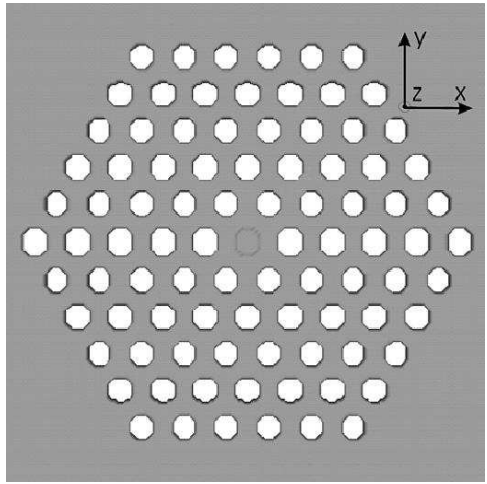, width=2.5in}
\caption{Deformation of the analyzed structure introduced by the application of mirror boundary conditions along the $x$
and $y$ axes and through the center of the defect. The applied mirror boundary conditions select the $x$-dipole mode, whose electric field pattern is shown in
Figure \ref{fig:xydn24}. Holes on the $x$-axis (in the central row, containing the defect) are elongated by 1 point in the $y$ direction. The refractive index of
the defect is $n_{defect}=2.4$. Parameters of the structure are $r/a=0.3$, $d/a=0.6$ and in the used discretization $a=15$.} \label{fig:add2n24}
\end{center}
\end{figure}

In our previous work, mirror boundary conditions were applied in the $x$, $y$ and $z$ directions to achieve an eight-fold reduction in the computational grid
size. We have since realized that the manner in which even (symmetric) mirror boundary conditions are implemented in our finite-difference time domain (FDTD)
code results in numerical output that properly corresponds to an {\em analyzed} structure with slight deformations relative to the {\em intended} structure. For
example, the set of mirror boundary conditions used to select the $x$-dipole mode in a defect cavity leads to a deformation of the structure as shown in Figure
\ref{fig:add2n24}. Holes on the $x$-axis (including the defect) are elongated in the $y$ direction by 1 point but the distance between holes in $x$ and $y$
directions is preserved. Because hole-to-hole distances are preserved under this deformation, the half-spaces $y>1/2$ and $y<-1/2$ actually maintain the
unperturbed PC geometry when holes in the central row are elongated by $1/2$ points in both the $\pm y$ directions. The symmetry of the PC surrounding the defect
is therefore broken, and this contributes to artificially high quality factors for $x$-dipole modes. An even mirror BC was also applied in the $z$ direction in
our previous analysis, causing a slight error in the thickness of the slab. The correct $d/a$ ratios of the structures analyzed in Ref.~\cite{ref:Painter98}
would be 0.6, 0.46 and 1, instead the values of 0.53, 0.4 and 0.93, as noted there.

\subsubsection{Perfectly symmetric cavity}

We re-analyzed the structure with $r/a=0.3$, $d/a=0.6$,
$n_{defect}=2.4$, the refractive index of the slab equal to
$n=3.4$, 5 layers of holes around the defect, and $a=15$ grid
points. For this set of parameters, the predicted $Q_\perp$ in
Ref.~\cite{ref:Painter98} for the $x$-dipole mode was $3\times
10^4$. In the present analysis, even mirror boundary conditions
were applied to the lower boundary in the $z$ direction only, to
reduce the computation size by one half and to eliminate TM-like
modes. Absorbing boundary conditions were applied to all
boundaries in $x$ and $y$ directions and to the upper boundary in
$z$ direction. Under these BC's the intended symmetry of the
defect structure was achieved. The initial field distribution was
chosen in such a way as to excite $x$ or $y$ dipole modes
selectively. For the $x$-dipole mode we now calculate
$Q_{||}=2260$, $Q_{\perp}=1730$ and $\frac{a}{\lambda}=0.3137$,
and for the $y$-dipole mode we calculate $Q_{||}=1867$,
$Q_{\perp}=1007$ and $\frac{a}{\lambda}=0.3182$.

\begin{figure}[hbt]
\begin{center}
\subfigure{\epsfig{figure=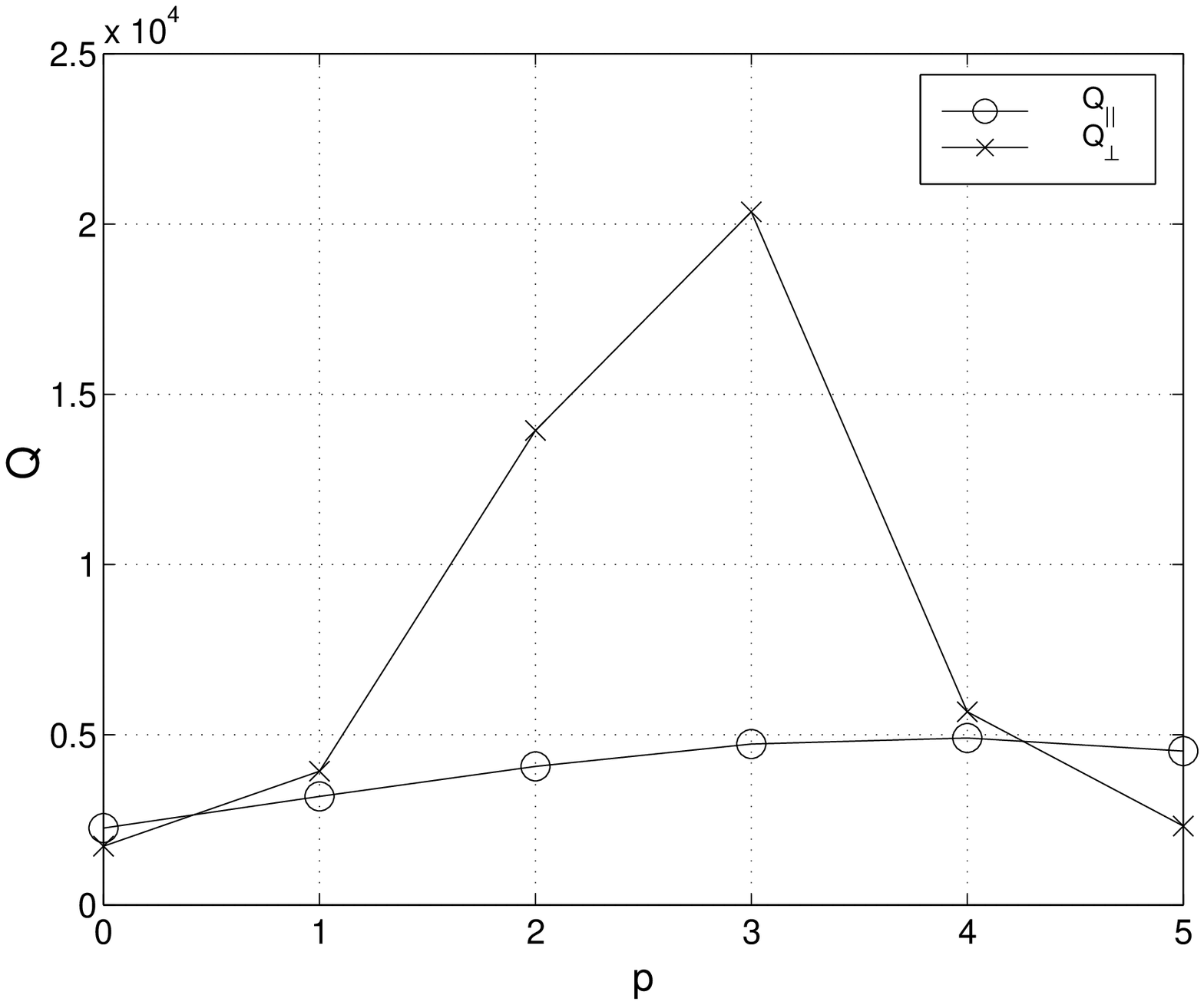, width=3in}}
\subfigure{\epsfig{figure=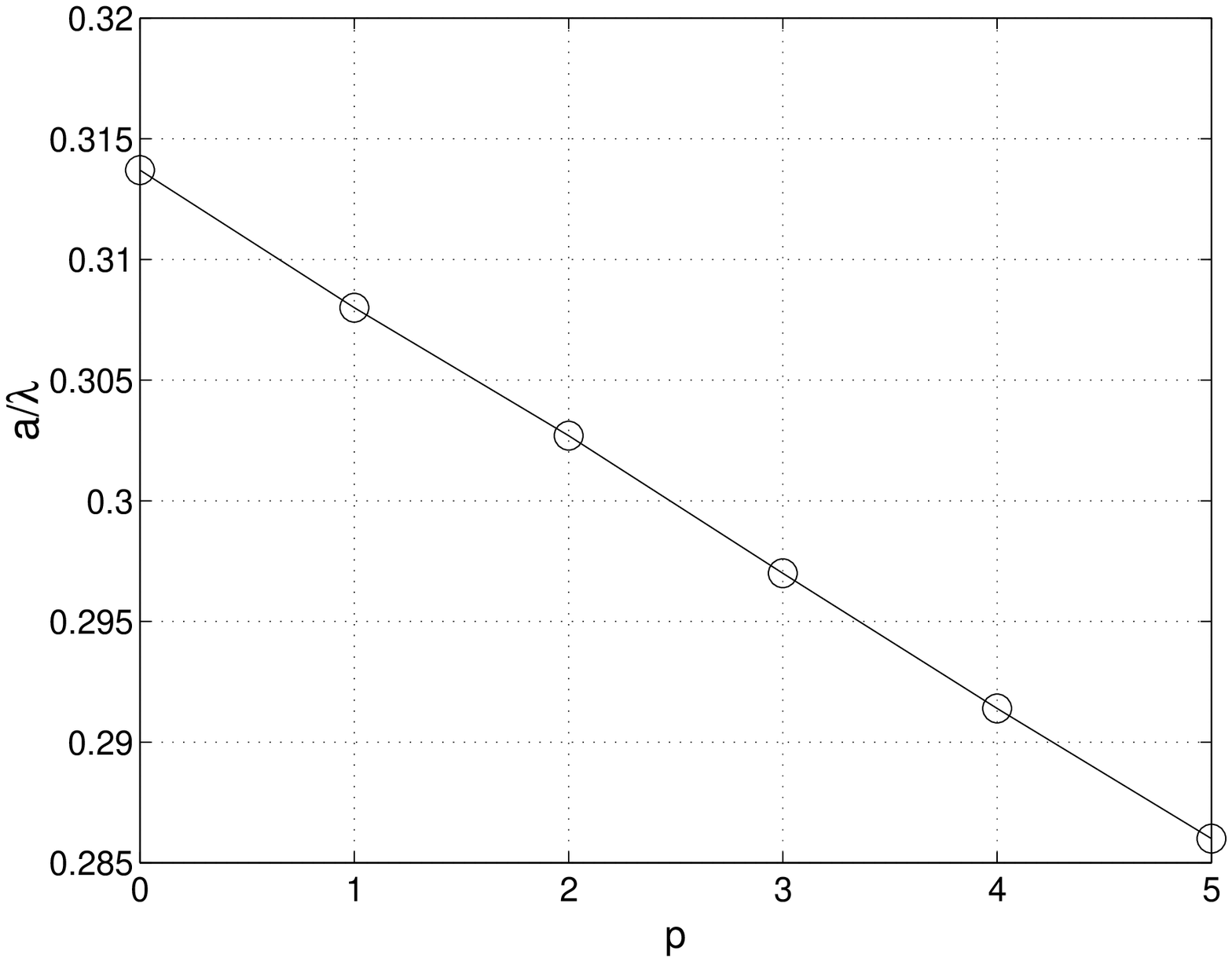, width=3in}}
\caption{(a) $Q$ factors and (b) frequencies of $x$-dipole modes in the structure shown
in Figure \ref{fig:add2n24}, as a function of the elongation parameter $p$. The structure parameters are $r/a=0.3$, $d/a=0.6$, $n=3.4$, $n_{defect}=2.4$, $a=15$
and 5 layers of holes surround the defect.} \label{fig:n24addp}
\end{center}
\end{figure}

The difference between parameters computed for the $x$ and $y$
dipole modes comes partly from the asymmetry of the structure
introduced by imperfect discretization. In a 2D PC with infinite
slab thickness, these two modes should be degenerate
\cite{ref:Painter98}. In the thin slab however, the $y$-dipole
mode suffers more vertical scattering at the edges of holes and,
therefore, has a lower $Q_\perp$. For a non-symmetric applied
initial field distribution leading to excitation of both $x$ and
$y$ dipole modes, we calculated $Q_{||}=2070$, $Q_{\perp}=1290$
and $\frac{a}{\lambda}=0.316$. This mode can be represented as a
superposition of the $x$ and $y$ dipole, with weighting factors
depending on the initial field. Its quality factor and resonant
frequency depend on these weighting factors.

\begin{figure}[hbt]
\begin{center}
\epsfig{figure=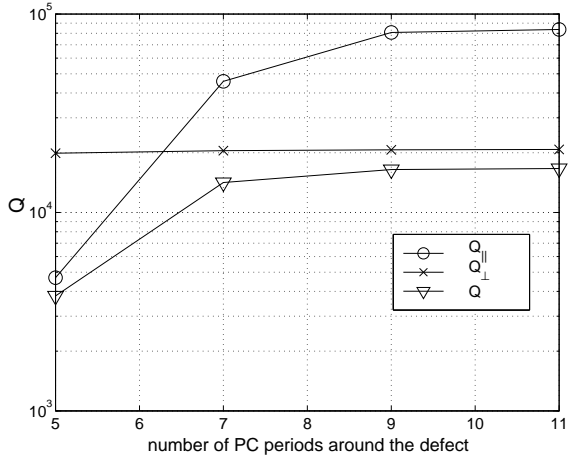, width=3in}
\caption{$Q$ factor for
$p=3$, in the structure shown in Figure \ref{fig:add2n24}, as a
function of number of PC periods around the defect. The structure
parameters are $r/a=0.3$, $d/a=0.6$, $n=3.4$, $n_{defect}=2.4$ and
$a=15$.} \label{fig:n24addp3}
\end{center}
\end{figure}

\subsubsection{Asymmetric cavity}

\label{subsection:n24addp} Our explanation for the discrepancy
between our current and previous results leads us to ask whether a
``real'' (intentional) elongation of the central row of holes
along the $y$-axis may actually improve the quality factor ($Q$)
of the $x$-dipole mode. A possible disadvantage of this approach
is the excitation of additional states caused by enlarging holes
in the central row. We have therefore analyzed how $a/\lambda$ and
$Q$ of the $x$-dipole mode in this structure changes as a function
of the elongation parameter $p$.  Holes on the $x$-axis (including
the defect) are elongated in both the $\pm y$ directions by $p/2$
points, in such a way that hole-to-hole distances are preserved
and the half-spaces $y>p/2$ and $y<-p/2$ maintain the unperturbed
PC geometry. The structure parameters are $r/a=0.3$, $d/a=0.6$,
$n=3.4$, $n_{defect}=2.4$ and 5 layers of holes surround the
defect. The periodicity $a$ used in these calculations is equal to
$15$ grid points, and the elongation step $\Delta p=1$ corresponds
to $a/15$. The results are shown in Figure \ref{fig:n24addp}. It
is interesting that the frequency of the mode decreases as $p$
increases, even though the amount of low refractive index material
increases. However, the net amount of low refractive index
material does not matter; what does matter is where the low
refractive index is positioned relative to the unperturbed PC. The
explanation of the decrease in frequency is very simple, if we
recall the $x$-dipole mode pattern shown in Figure
\ref{fig:xydn24}. This is a donor type defect mode, which
concentrates its electric field energy density in low refractive
index regions of the unperturbed PC. As $p$ increases, layers of
PC holes are moved away from the defect in $y$ direction. For
example, the $n$-th layer of holes parallel to the $x$-axis will
be positioned at $y=\pm n a \sqrt{3}/2 \pm p/2$, instead of $y=\pm
n a \sqrt{3}/2$. Therefore, a large refractive index material will
be positioned at places where the mode expects to ``see'' air,
leading to a decrease in mode's frequency. By tuning the mode's
frequency across the band-gap, we can also tune its $Q$ factor, as
noted previously \cite{ref:JV_spie2001}. For $p=3$, $Q_\perp$
reaches the value of $20000$. The analyzed $x$-dipole mode mostly
resonates in the direction of $y$-axis, i.e. in the $\Gamma X$
direction of photonic crystal. The tuning of the elongation
parameter $p$ is, therefore, analogous to tuning of a spacer in
the micropost cavity, which leads to tuning of mode's frequency
and $Q$ factor. A more detailed explanation of the effect of
elongations on $Q$ factors is given in the appendix of this
article. By increasing the number of PC periods around the defect,
the total quality factor $Q$ approaches $Q_\perp$, as shown in
Figure \ref{fig:n24addp3}. It is important to note that $Q_{||}$
does not increase exponentially with the number of PC layers
around the defect. Instead, it saturates at large number of PC
layers. The reason is in the choice of a boundary for separation
of $Q_{\perp}$ from $Q_{||}$, positioned at approximately
$\lambda/2$ from the surface of the membrane.
\begin{figure}[hbtp]
\begin{center}
\epsfig{figure=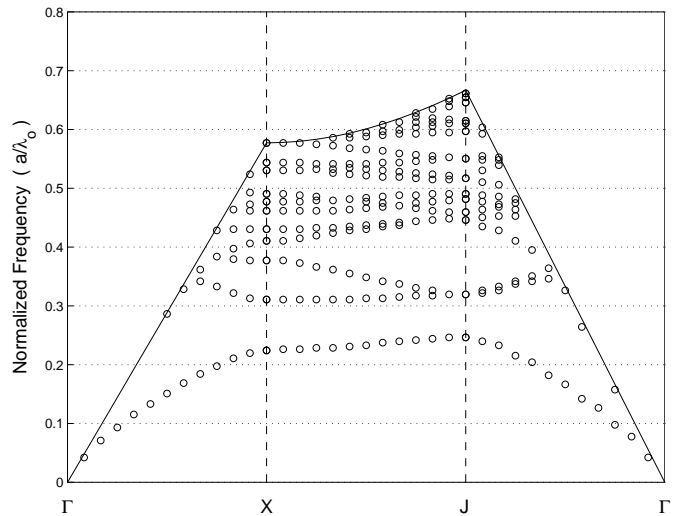, width=3.5in}
\caption{Band diagram for TE-like modes of a thin slab ($n=3.4$, $d/a=0.75$) surrounded by air on both sides and patterned
with a hexagonal array of air holes ($r/a=0.275$).} \label{fig:bandra275_da075}
\end{center}
\end{figure}
 From the radiation pattern of the $x$-dipole mode, we can see
that some portion of the out-of-plane loss (mostly in the $x$
direction) still gets collected in $Q_{||}$. This loss cannot be
suppressed by increasing the number of PC layers around the
defect, and it determines the value at which $Q_{||}$ saturates.
However, much larger out-of-plane loss is collected in $Q_\perp$, which ultimately determines the total quality factor $Q$.\\
Therefore, a dramatic improvement in $Q$ factors of dipole modes over single defect microcavities can be obtained by introducing a novel type of PC lattice defect, consisting of the elongation of holes along the symmetry axes. We call it a {\it fractional edge dislocation}, by analogy with edge dislocations in solid state physics. Edge dislocations are formed by introducing extra atomic planes into the crystal lattice. On the other hand, we here insert only fractions of atomic planes along the symmetry axes of photonic crystal.\\
\indent Unfortunately, at this time, we do not know how to control
the refractive index of a single PC hole during the fabrication.
For this reason, in the next subsection we will consider
alternative methods of forming single defect microcavities, which
are much easier to construct by microfabrication.

Even after the elongation of holes on the $x$-axis by 1 point, $Q$
factors of $30000$ were not obtained. This means that the
application of mirror BC at $x=0$ and $y=0$ planes causes
additional effects that lead to the overestimation of quality
factors. One of the reasons may be that the excited dipole mode
does not have a symmetry described by the applied discretized
mirror BC's. This may be partly due to the structure imperfection
caused by discretization. In order to avoid problems caused by
BC's all the calculations in this paper were done by applying the
even mirror symmetry to $z=0$ plane only, in order to select
TE-like modes and to reduce the computation size by one half.
Absorbing boundary conditions were applied to all boundaries in
$x$ and $y$ directions and to the upper boundary in the $z$
direction. To prove that the application of mirror BC at the lower
$z$ boundary does not change $Q$, we also analyzed entire
structures with absorbing BC's applied to all boundaries and
obtained the same results as in the analysis of one half of the
structure.

\begin{table}
\begin{center}
\caption{$Q$ factors of dipole modes excited in microcavities
formed by decreasing the radius of a single PC hole.}
\vspace{0.5cm}
\begin{tabular}{|c|c|c|c|c|c|}
\hline
$r/a$ & $r_{def}/a$ & $d/a$ & $a/\lambda$ & $Q_{||}$ & $Q_\perp$\\
\hline
    0.275 & 0.15 & 0.75 & 0.286 & 778 & 920\\
\hline
    0.275 & 0.2 & 0.75 & 0.297 & 470 & 2078\\
\hline
    0.25 & 0.15 & 0.75 & 0.277 & 230 & 1840\\
\hline
    0.25 & 0.2 & 0.75 & 0.284 & 116 & 3190\\

\hline
\end{tabular}
\label{table:Qsdd}
\end{center}
\end{table}

\subsection{Reducing the radius of a single hole}

Microcavities analyzed in this subsection are formed by reducing
the radius of a single hole to $r_{def}$. We calculated parameters
of excited dipole modes for a range of microcavity parameters and
results are shown in Table \ref{table:Qsdd}. For all tabulated
results, 5 layers of holes surround the defect and $a=20$.
Calculated $Q$'s are not very impressive, but they do provide us
with a good starting point for further optimization. Why did we
decide to use a relatively small $r/a$ ratio for the PC?
According to our calculations, increasing $r/a$ within the analyzed range leads to increases in the band gap and reduction of lateral losses. However, vertical
scattering at the edges of holes also increases and $Q_{\perp}$ drops. It is therefore important to find an optimum $r/a$ which leads to small vertical losses
but preserves good lateral confinement, in order not to increase the mode volume too much.

The band diagram for TE-like modes of a thin slab ($n=3.4$,
$d/a=0.75$) surrounded by air on both sides and patterned with a
hexagonal array of air holes ($r/a=0.275$) is shown in Figure
\ref{fig:bandra275_da075}. These PC parameters are used in most of
the calculations in the next section. From the comparison of the
dipole mode frequencies tabulated in Table \ref{table:Qsdd} and
the band diagram shown in Figure \ref{fig:bandra275_da075}, it can
be confirmed that as the cavity mode's frequency approaches the
bottom of the air band, vertical losses decrease, but lateral
losses increase, which in turn leads to an increase in the mode
volume. Our goal in the next section will be to reduce vertical
losses and improve $Q$ factors even further, while preserving
small mode volumes. For this purpose we will explore fractional
edge dislocations. In Figure \ref{fig:n24addp}, one can observe
that an increase in the elongation parameter $p$ can be used to
tune the $Q_\perp$ factor of a mode, but also leads to a decrease
in the dipole mode's frequency. This implies that by increasing
$p$, the mode is pulled deeper into the band gap, away from the
air band edge, which leads to its better lateral confinement .
Therefore, we can simultaneously achieve a reduction in vertical
losses and an improvement in lateral confinement (i.e. an increase
in both $Q_\perp$ and $Q_{||}$, and a reduction in the mode
volume).

\section{Cavities for strong coupling}

In this section we consider the design of PC microcavities to
achieve strong coupling between the cavity field and a single
gas-phase atom, that is, an atom located in free space rather than
contained as an impurity in the dielectric slab. Our long-term
goal is to investigate photonic bandgap structures for single-atom
cavity quantum electrodynamics in the strong coupling regime
\cite{ref:Kimb94}.  For this purpose, the microcavity mode quality
factor ($Q$) has to be as large as possible and the mode volume
($V_{mode}$) as small as possible. These two design rules are also
followed when designing PC microcavities for semiconductor lasers.
However, in a cavity for strong coupling, an atom must be trapped
at the point where it interacts most strongly with the cavity
field. Therefore, an additional design goal is imposed in this
case: the cavity mode should have the E-field intensity as high as
possible in the air region. When designing a laser cavity, the
problem is opposite: one tends to maximize the overlap between the
gain region and the cavity field and, therefore, wants to have the
strongest E-field in the semiconductor region.

\begin{figure}[t]
\begin{center}
\subfigure{\epsfig{figure=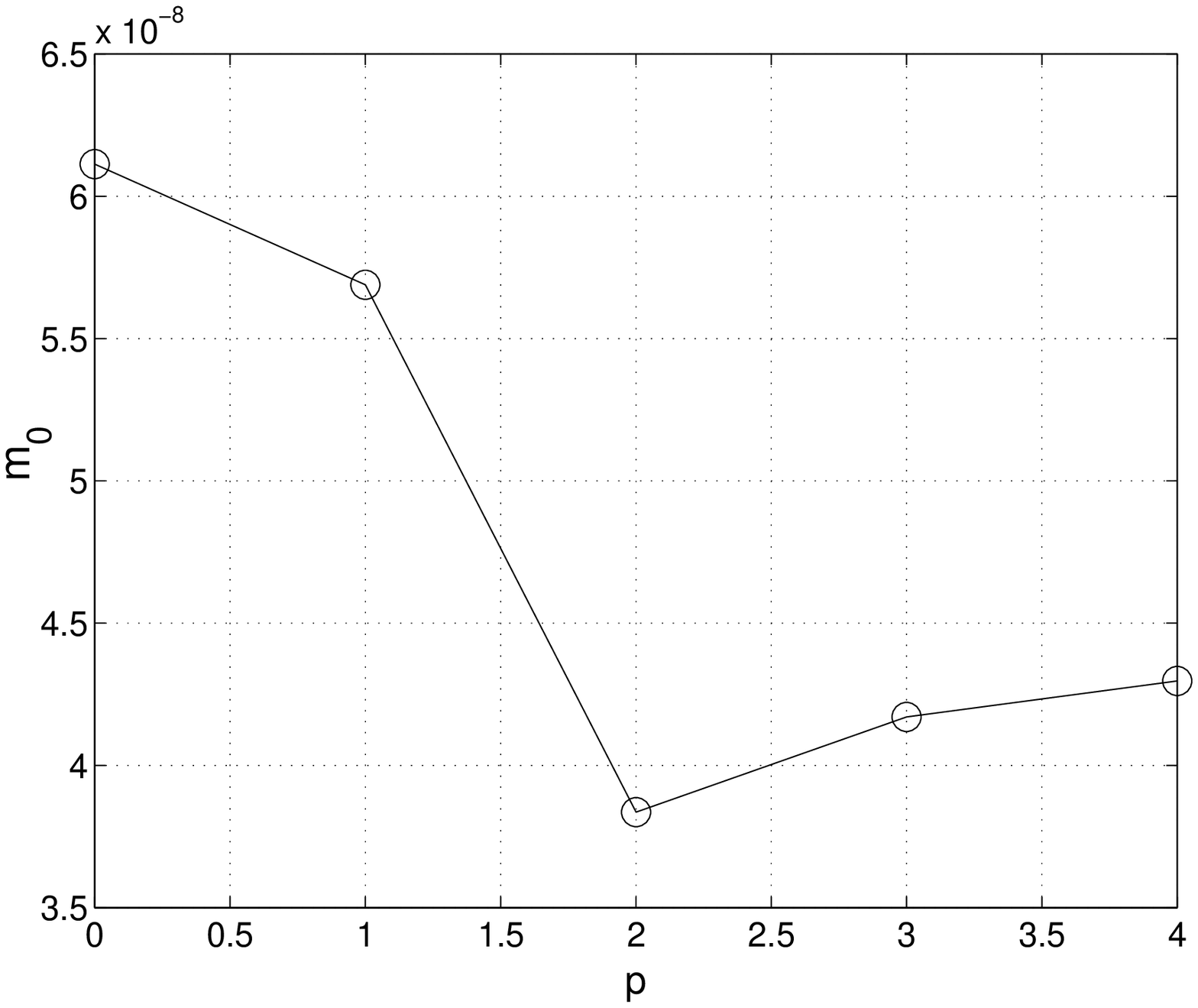, width=3in}}
\subfigure{\epsfig{figure=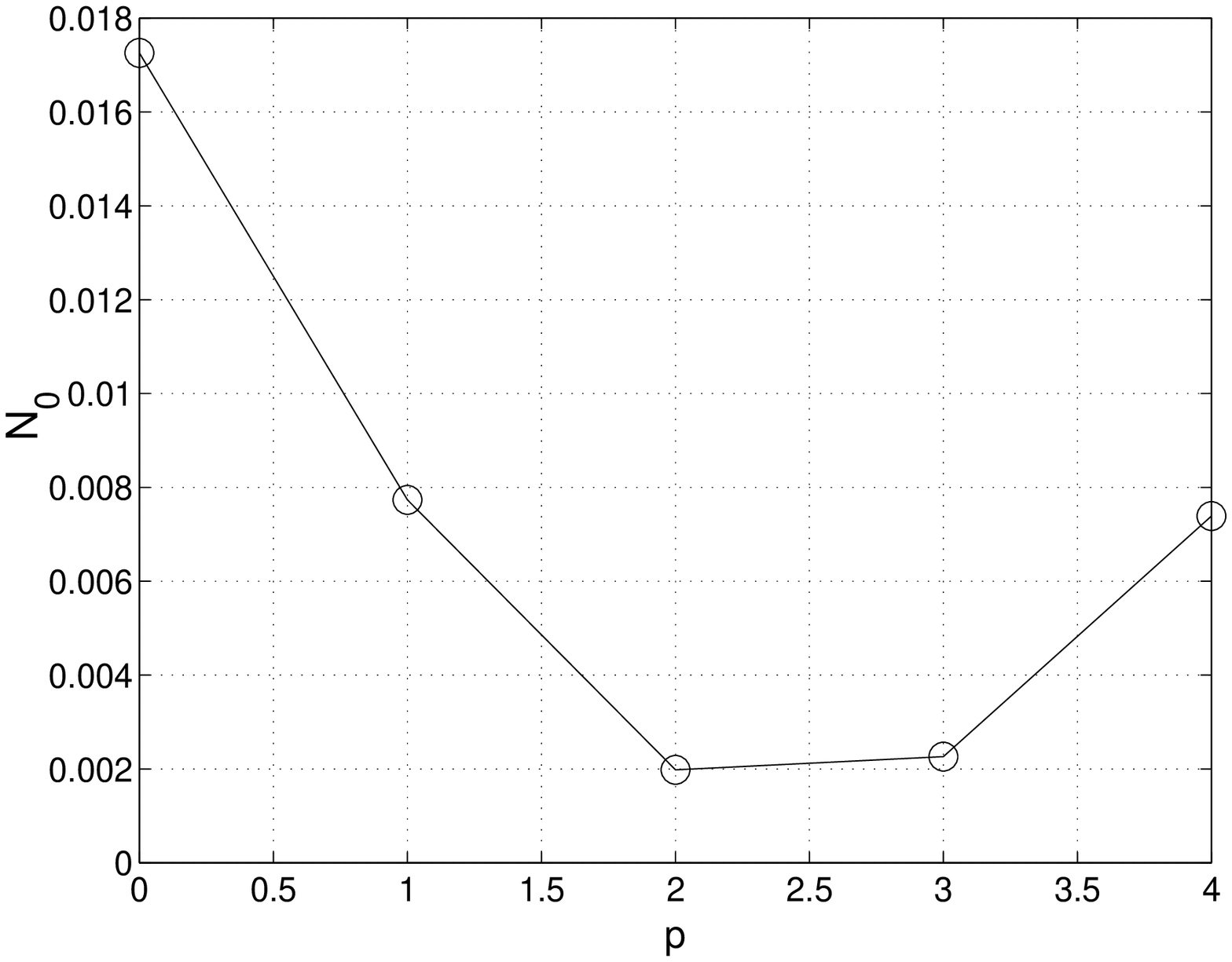, width=3in}}
 \caption{Parameters of the $x$-dipole mode in a single defect structure ($r/a=0.275$,
$d/a=0.75$, $r_{def}/a=0.2$, $n=3.4$ and $a=20$) as a function of the elongation parameter $p$: (a) $m_0$; (b) $N_0$} \label{fig:addp2}
\end{center}
\end{figure}

Mode volume ($V_{mode}$), critical atom ($N_0$) and photon ($m_0$) numbers are defined as follows:

\begin{eqnarray}
V_{mode}&=\frac{\iiint \epsilon(r)|E|^2 dV}{max[\epsilon(r)|E|^2]}\\
N_0&=\frac{2\kappa \gamma_\perp}{g^2}\\
m_0&=\biggl ( \frac{\gamma_{\perp}}{2g}\biggr )^2,
\end{eqnarray}
where $\kappa$ is the cavity field decay rate, proportional to the
ratio of the angular frequency of the mode ($\omega_0$) and the
mode quality factor ($Q$):
\begin{equation}
\kappa=\frac{\omega_0}{4\pi Q}
\end{equation}
$\gamma_\perp$ is the atomic dipole decay rate ($2.6$ MHz for
Cesium) and $g$ is the coupling parameter at the point where we
want to put an atom:
\begin{equation}
g(r)=g_0\frac{ \epsilon(r)|E|}{max[\epsilon(r)|E|]}
\end{equation}
$g_0$ denotes the vacuum Rabi frequency:
\begin{eqnarray}
g_0&=\gamma_\perp \sqrt{\frac{V_0}{V_{mode}}}\\
V_0&=\frac{c\lambda^2}{8\pi\gamma_\perp}
\end{eqnarray}

Strong coupling is possible if both $N_0$ and $m_0$ are smaller
than 1. Therefore, in order to predict whether the strong coupling
can occur, we must calculate upper limits of $N_0$ and $m_0$ and
compare them to 1. In other words, it is acceptable if calculated
critical numbers are overestimated. As the number of PC layers
around the defect increases, the total quality factor $Q$
approaches $Q_{\perp}$ and $V_{mode}$ drops due to the better
lateral confinement. Hence, we can calculate $N_0$ and $m_0$ by
assuming $Q=Q_{\perp}$
and using $V_{mode}$ calculated for 5 PC layers around the defect. \\
For all calculations in this section, the refractive index of the
slab is $n=3.4$, 5 layers of holes surround the central hole and
$a=20$. The elongation step $\Delta p=1$ corresponds to $a/20$,
i.e. $5\%$ of the lattice periodicity $a$. The material and PC
properties are chosen in such a way that cavities operate at
$\lambda=852$ nm (the wavelength corresponding to the D2 atomic
transition in $^{133}Cs$).

\subsection{Single defect with fractional edge dislocations}

\label{subsection:singledef_fed}

Let us study microcavities formed by reducing the radius of a
single hole and simultaneously applying a fractional edge
dislocation (of order $p$) along the $x$-axis. It is important to
note that all holes along the $x$-axis are elongated in this
process, including the defect hole of the reduced radius. We will
calculate the dependence of the $x$-dipole mode properties on
parameter $p$. Parameters of the unperturbed PC are: $r/a=0.275$,
$d/a=0.75$, $a=20$ and the defect hole radius is $r_{def}/a=0.2$.
The critical atom ($N_0$) and photon ($m_0$) numbers are shown as
a function of the elongation parameter $p$ for the $x$-dipole mode
in Figure
\ref{fig:addp2}. $V_{mode}$ does not change significantly with $p$
and it is approximately equal to $0.1(\lambda/2)^3$ for all
structures. When $p$ increases, the frequency of the mode moves
away from the band edge, towards the center of the band gap,
$Q_{||}$ increases and $Q_\perp$ peaks at the value of
$1\times10^4$ for $p=2$.

\begin{figure}[hbtp]
\begin{center}
\subfigure{\epsfig{figure=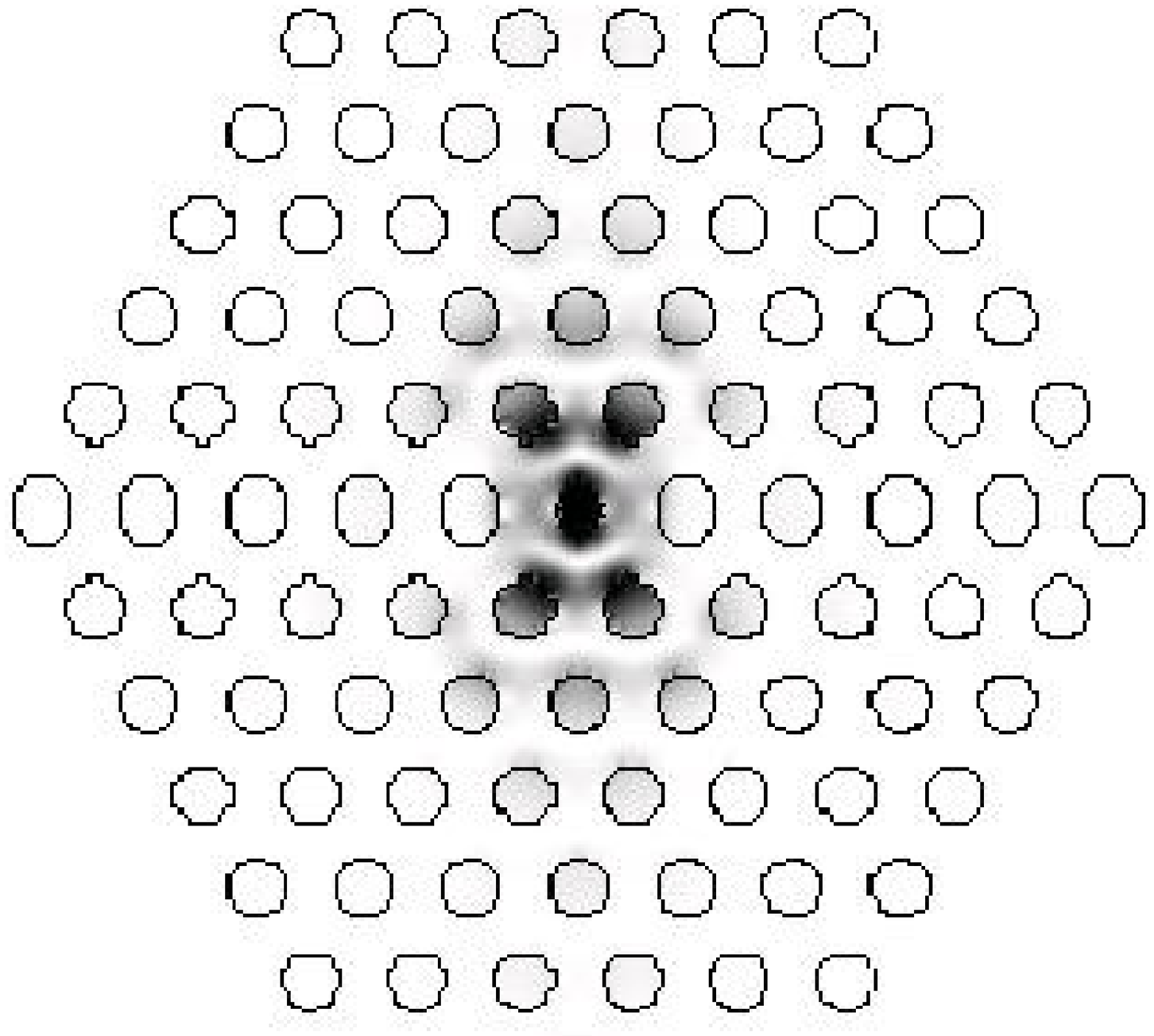, width=2.9in}}
\subfigure{\epsfig{figure=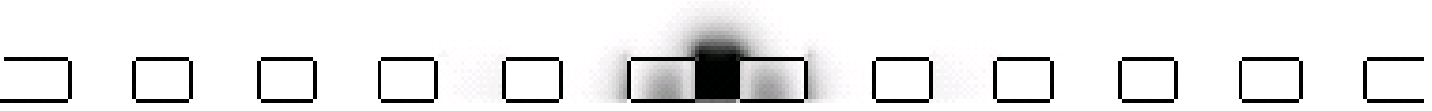, width=3in}}
\caption{Electric field intensity pattern of the $x$-dipole mode excited in a
microcavity formed by reducing the radius of a single hole and simultaneously elongating holes on the $x$-axis by 2 points. (a) slice through the middle of the
membrane, in the $x-y$ plane; (b) $x-z$ half-plane ($z>0$).} \label{fig:Eadd2}
\end{center}
\end{figure}

From the electric field intensity pattern of the $x$-dipole mode
shown in Figure \ref{fig:Eadd2}, one can see that the electric
field intensity is very strong within the defect hole. Therefore,
an atom trapped there should interact very strongly with the
cavity field. From the calculated critical atom and photon
numbers, it then should be possible to achieve very strong
coupling. At $\lambda=852nm$, the parameters of such a cavity are
$r=70nm$, $d=190nm$, $a=250$ and $r_{def}=50nm$. Due to extremely
small mode volumes in these cavities, strong coupling is possible
even for moderate values of $Q$, as $Q_\perp$ did not exceed $\sim
1\times 10^4$ in the parameter range of Fig.~\ref{fig:addp2}.
Furthermore, $m_0$ is much smaller than $N_0$, which means that we
can try to improve $Q$ factors further at the expense of
increasing $V_{mode}$. $Q$ factors above $1\times10^4$ and similar
values of $m_0$ and $N_0$ can also be obtained for the cavity
consisting of a single defect with $r_{def}/a=0.2$ and a
fractional edge dislocation of order $p$, produced in the photonic
crystal whose parameters are $r/a=0.3$, $d/a=0.65$ and $n=3.4$.

\begin{figure}[hbtp]
\begin{center}
\epsfig{figure=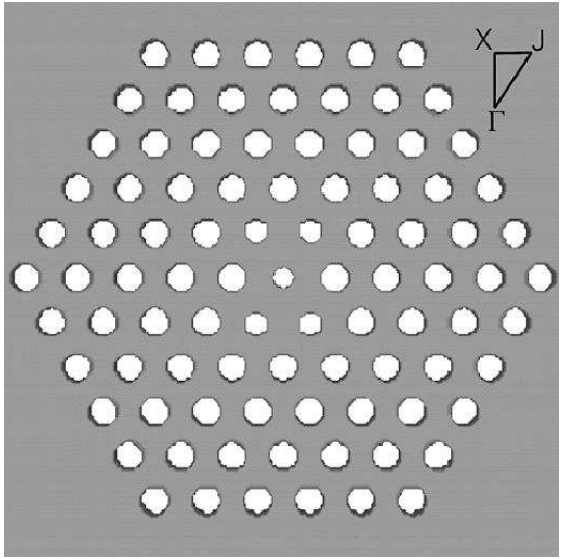, width=2.9in}
\caption{Tuning 4 holes
closest to the defect in $\Gamma J$ directions. Their radii are
reduced to $r_1$ and they are simultaneously moved away from the
defect in $\Gamma J$ directions by $r-r_1$. The radius of the
central hole is $r_2$.} \label{fig:struc4holes}
\end{center}
\end{figure}

\subsection{Tuning holes around the defect}

\begin{figure}[hbtp]
\begin{center}
\epsfig{figure=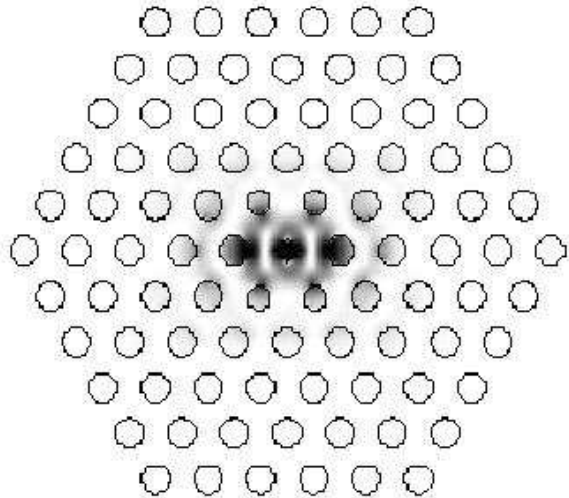, width=3in}
\caption{Electric field intensity pattern of the $y$-dipole mode excited in the cavity where 4 holes closest to the defect
in $\Gamma J$ directions are tuned. Their radii are reduced to $r_1/a=0.225$ and they are simultaneously moved away from the defect in $\Gamma J$ directions by
$r-r_1$. The radius of the central hole is $r_2/a=0.2$. PC parameters are $r/a=0.275$, $d/a=0.75$, $a=20$ and $n=3.4$.} \label{fig:yd4holes}
\end{center}
\end{figure}

Dipole modes are particularly sensitive to the geometry of holes
closest to the defect. By tuning these holes, we can induce
frequency splitting of dipole modes and dramatically influence
their $Q$ factors. In Ref.~\cite{ref:Painter98}, the variation of
two nearest neighbor holes along the $x$-axis was analyzed. Here,
we will test the influence of changing 4 holes closest to the
defect in $\Gamma J$ directions. The analyzed structure is shown
in Figure \ref{fig:struc4holes}. The radius of the central hole is
reduced to $r_2$ and the radii of the 4 closest holes in $\Gamma
J$ directions are reduced to $r_1$. These 4 holes are
simultaneously moved away from defect, by $r-r_1$ in $\Gamma J$
directions, which preserves the distance between them and the next
nearest neighbors in the same directions. This design will improve
the $Q$ factor of the $y$-dipole mode and spoil the $Q$ of the
$x$-dipole mode. We analyzed structures with various parameters,
but our best result was obtained for $r/a=0.275$, $d/a=0.75$,
$r_2/a=0.2$, $r_1/a=0.225$ and $a=20$. The electric field
intensity pattern of the excited $y$-dipole mode is shown in
Figure \ref{fig:yd4holes}, and its calculated parameters are
$a/\lambda=0.289$ and $Q_\perp=4890$. From Table \ref{table:Qsdd}
we can see that a dipole mode excited in a single defect
microcavity with this $r_2/a$, $r/a$, $d/a$ has $Q_{\perp}=2078$.
Therefore, the tuning of 4 holes can lead to a substantial
increase in $Q$ of the $x$-dipole mode. The disadvantages of this
design include the excitation of defect modes other than dipoles
(coming from variation of several holes).

\begin{figure}[hbtp]
\begin{center}
\epsfig{figure=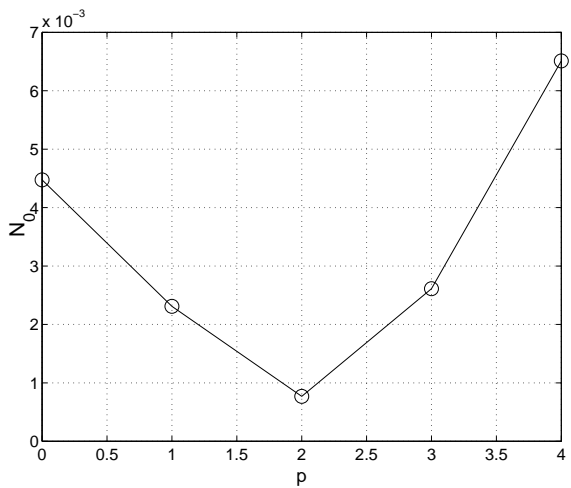, width=3in}
\caption{$N_0$ of the $x$-dipole mode in the structure where 4 holes in $\Gamma J$ directions are tuned (as shown in Figures
\ref{fig:struc4holes} and \ref{fig:yd4holes}), as a function of the elongation parameter $p$. Holes on the $y$-axis are elongated by $p$ points in the $x$
direction in such a way that the half-spaces $x>p/2$ and $x<-p/2$ remain the unperturbed PC geometry.} \label{fig:addp4}
\end{center}
\end{figure}

Let us try to improve $Q$ of this cavity even further, by also
employing the idea of elongation of holes along desired
directions. Our mode of choice is the $y$-dipole and we will
elongate holes sitting on the $y$-axis by $p$ points in the $x$
direction in such a way that the half-spaces $x>p/2$ and $x<-p/2$
remain the unperturbed PC geometry.
The dependence of
$N_0$ (which decreases with $Q$) on parameter $p$ is shown in
Fig.~\ref{fig:addp4}. $V_{mode}$ does not change significantly
with $p$ and is in the range between $0.09(\lambda/2)^3$ and
$0.12(\lambda/2)^3$. The calculated $m_0$ is around $5 \cdot
10^{-8}$ for all structures. Again, a very strong coupling is
achievable by this design, and we note that $Q_{\perp}$ at the
point $p=2$ achieves a value $\approx 3.3\times 10^4$.

\subsection{Atomic physics}

Given that microcavities with strong coupling parameters can be
designed and fabricated, two further technical issues must be
addressed in order to establish the feasibility of cavity QED with
neutral atoms in PC's.  First, we must identify a method for
stably trapping an individual atom within a hole of the PC.
Second, accurate estimates or measurements must be made of the
surface interaction between such a trapped atom and the
semiconductor substrate. Although definitive solutions to these
challenges are the subject of future work, we include in this
subsection a brief discussion of each topic.

In addition to the creation of photonic crystals, modern micro-fabrication techniques enable the patterning of either conductive wires or ferromagnetic materials
at the micron scale and below. As a result, it should be possible to construct magnetic microtraps with field curvatures $\sim 10^8$ G/cm$^2$
\cite{ref:Weinstein95,ref:Drndic01,ref:Haensch,ref:Juerg}. With this magnitude of field curvature an Ioffe trap could hold a Cs atom with Lamb-Dicke parameter
$\eta\sim 0.035$ in the radial direction, yielding $\Delta x\sim 10$ nm in the ground state of the trapping potential and enabling resolved-sideband laser
cooling as a means of putting single atoms in the ground state.

We are currently investigating fabrication and laser cooling techniques for an atom-trapping scheme in which micron-scale wires would be deposited on the surface
of the PC semiconductor substrate, such that the circular wire pattern of an Ioffe microtrap are arranged concentrically around a defect microcavity. The trap
designs discussed in \cite{ref:Weinstein95} have sufficiently large inner diameter not to disturb the photonic bandgap structure of the defect cavity. Such a
wire arrangement would project a magnetic field with a stable minimum at the geometric center of the microcavity, such that one or more atoms could be confined
within the defect hole and would therefore experience strong coupling. A similar geometry could be envisioned for microtraps based on permanent magnets rather
than current-carrying wires, which would have significant advantages in terms of heat load to the PC substrate.

A neutral atom trapped within a hole of a PC structure will experience surface (van der Waals) interactions that are quite difficult to estimate. The
significance of such interactions is twofold. The sensitive dependence of the ground state energy shift on an atom's position will lead to mechanical forces that
must be compensated by the trap design. Differential shifting of the atomic ground and excited states on the cavity QED transition must also be accounted for, as
this will introduce a position-dependent detuning relative to the fixed microcavity resonance. Although van der Waals shifts can be computed for alkali atoms
near dielectric or metallic boundaries with simple symmetry \cite{ref:Barton97,ref:Chevrollier92}, the case of an atom in a PC hole is far more complex. The
local geometry seen by a trapped atom will be that of a cylindrical hole with finite extent, and a proper calculation must take into account the overall
modification of vacuum modes due to the extended photonic crystal.

\begin{figure}[htbp]
\begin{center}
\subfigure{\epsfig{figure=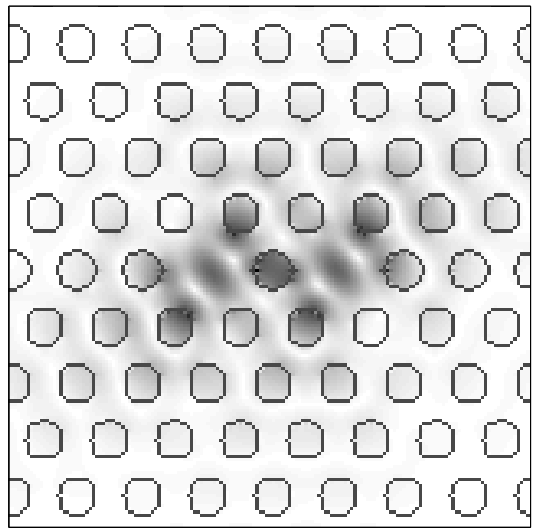, width=2.8in}}
\subfigure{\epsfig{figure=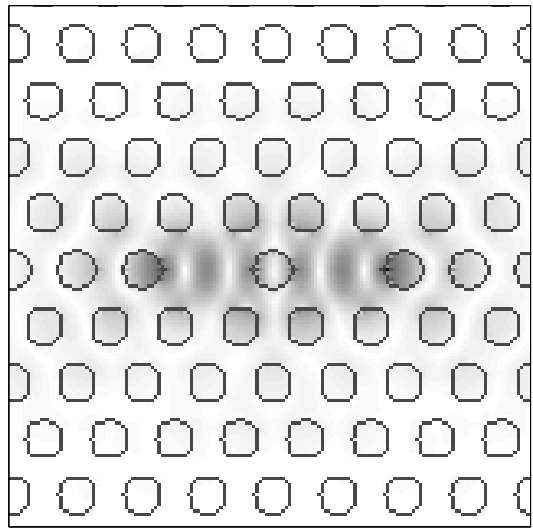, width=2.8in}}
\caption{Electric field intensity patterns of the coupled dipole modes: (a)
constructively; (b) destructively coupled defect states.} \label{fig:coupling}
\end{center}
\end{figure}

Furthermore, the resonant frequencies of many atomic transitions that connect to low-lying states, and therefore contribute strongly to their van der Waals
shifts, are above the bandgap of the semiconductor substrate.  Such transitions will ``see'' an absorptive surface while those below the bandgap will see a
dielectric surface. This set of factors brings the complexity of the desired calculation well beyond that of existing analytic results in the literature. It
should be noted that experimental measurements of the van der Waals shifts in our proposed system would be of significant interest for the general subject of
quantum electrodynamics of semiconductors.

We are pursuing a numerical strategy for estimating the magnitude
of surface interactions. In a linear response approximation, it
should be possible to compute leading-order contributions to the
van der Waals shifts from FDTD simulations of the electromagnetic
field created by an oscillating dipole source in the photonic
crystal structure. While the nature of the code does not allow us
to compute directly the backaction of the scattered field on the
source dipole, we believe that elementary field theory can be used
to relate the simulated field to van der Waals shifts. Our
findings will be reported in a forthcoming publication.

\begin{figure}[hbtp]
\begin{center}
\epsfig{figure=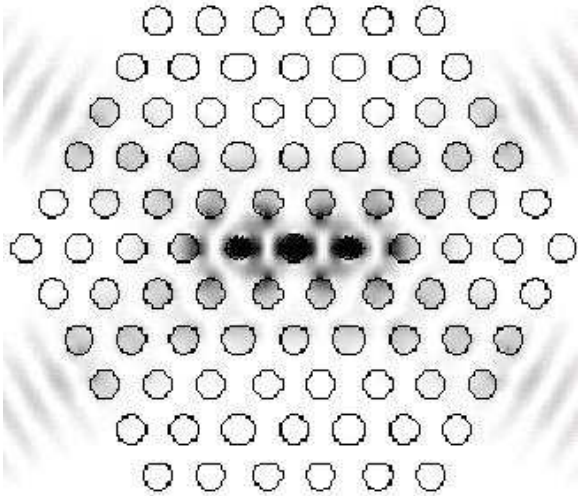, width=3in}
\caption{Electric field intensity patterns of the constructively coupled dipole modes in the structure with the following
parameters: $r_{def}/a=0.2$, $d/a=0.75$, $r/a=0.275$, $n=3.4$ and $a=20$. Holes in columns containing defects are elongated by 2 points in the $x$ direction}
\label{fig:spregxdips}
\end{center}
\end{figure}

\subsection{Coupled dipole defect modes}

The significance of surface effects that could perturb atomic radiative structure within the small defect hole is still unknown. For that reason, we will try to
investigate ways of increasing the radius of the hole where the coupling between the atom and the cavity field should occur. Let us now analyze the cavity design
where the strong E-field intensity can be achieved in the center of an unperturbed hole. The idea is to use coupling of two dipole defect states. Resonant modes
of the microcavity formed by coupling two single defects are presented in Figure \ref{fig:coupling}. Based on the resultant electric field intensity in the
central, unperturbed hole, we call them constructively or destructively coupled defect states. They have different frequencies, as well as $Q$ factors. We will
analyze properties of the constructively coupled state, since the central, unperturbed hole would be a good place for an atom.

\begin{figure}[hbtp]
\begin{center}
\epsfig{figure=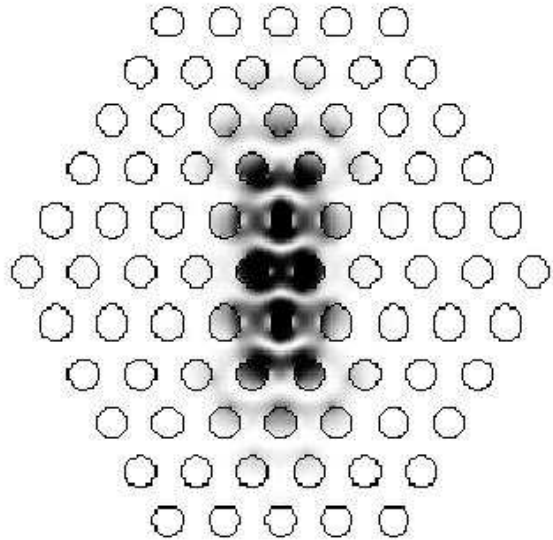, width=3in}
\caption{Electric field intensity patterns of the constructively coupled dipole modes in the structure with the following
parameters: $r_{def}/a=0.2$, $d/a=0.75$, $r/a=0.275$, $n=3.4$ and $a=20$. Holes in rows containing defects are elongated by 2 points in the $y$ direction.}
\label{fig:spregydips}
\end{center}
\end{figure}

We analyzed a series of structures with different parameters. The
best results were obtained for two coupled defects with
$r_{def}/a=0.2$ in a PC with the following parameters: $d/a=0.75$,
$r/a=0.275$, $n=3.4$ and $a=20$. Holes in the $\Gamma X$
direction, in columns containing defects, are elongated by 2
points in the $x$ direction. The mode pattern of the
constructively coupled defect state is shown in Figure
\ref{fig:spregxdips}. Parameters of the mode are:
$a/\lambda=0.29$, $Q_{\perp}=6100$, $V_{mode}=0.19(\lambda/2)^3$,
$m_0=1.5\cdot 10^{-7}$ and $N_0=0.0135$. An atom can now be
trapped in the central hole of unperturbed radius. For
$\lambda=852nm$, this radius is $r=68nm$, which is a significant
improvement over the previous design, where an atom must be
trapped within a $50nm$ radius hole. Again, a strong coupling is
achievable in this cavity.

An alternative way of forming the coupled defects state is represented in Figure \ref{fig:spregydips}. We use the same PC parameters as previously:
$r_{def}/a=0.2$, $d/a=0.75$, $r/a=0.275$, $n=3.4$ and $a=20$. Holes in rows containing defects, are elongated by 2 points in the $y$ direction.

\begin{figure}[hbtp]
\begin{center}
\subfigure{\epsfig{figure=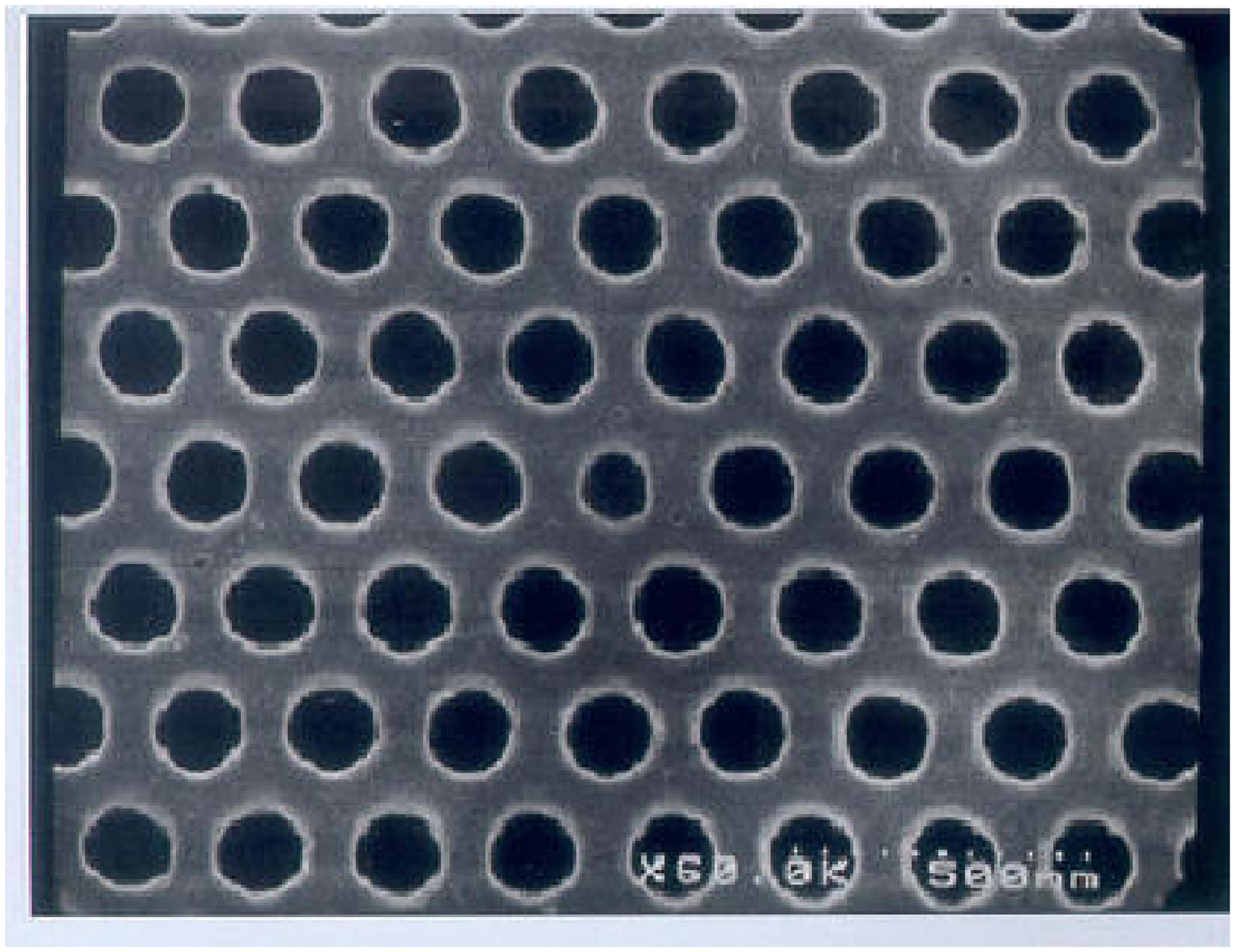, width=2.4in}}
\subfigure{\epsfig{figure=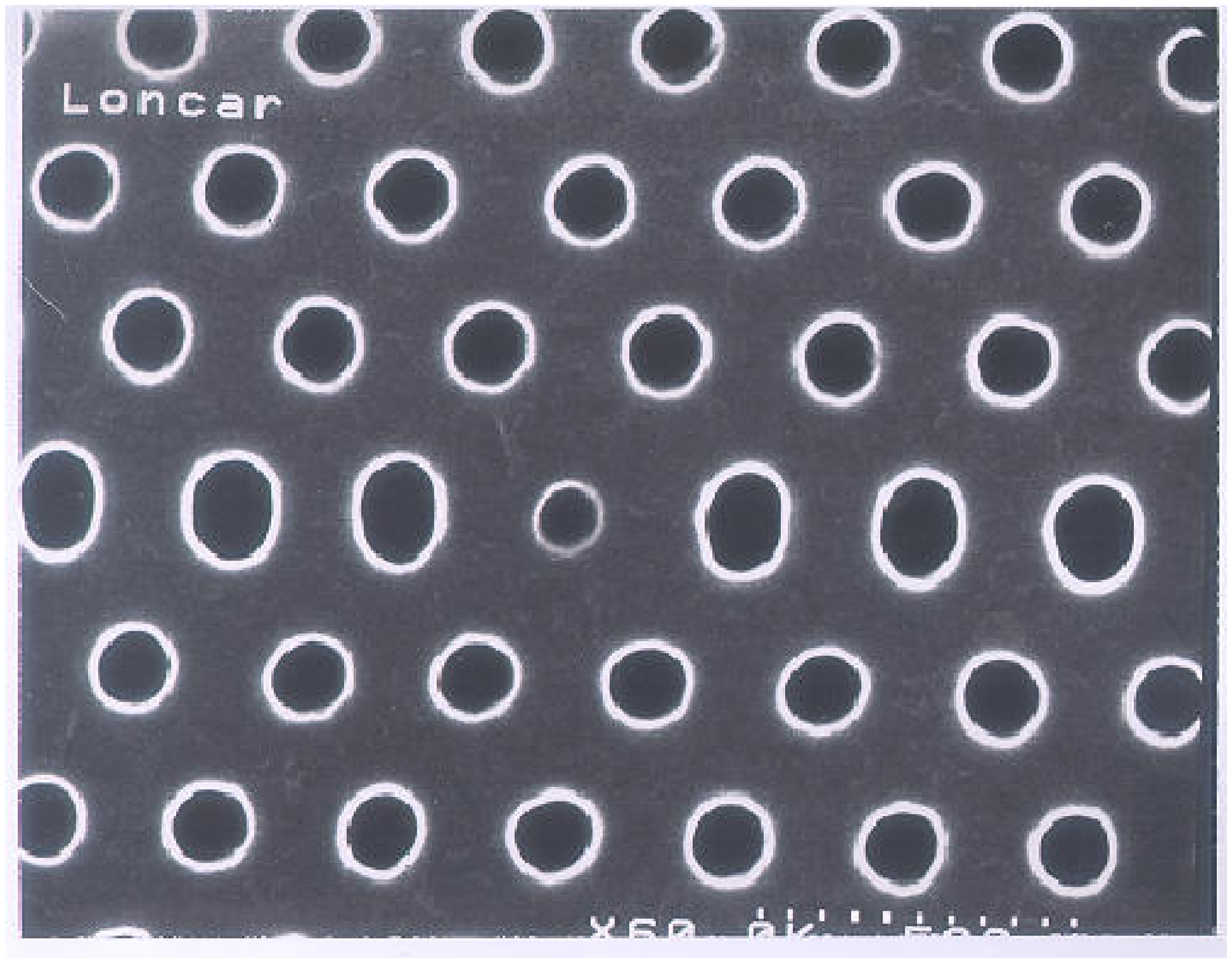, width=2.4in}}
\subfigure{\epsfig{figure=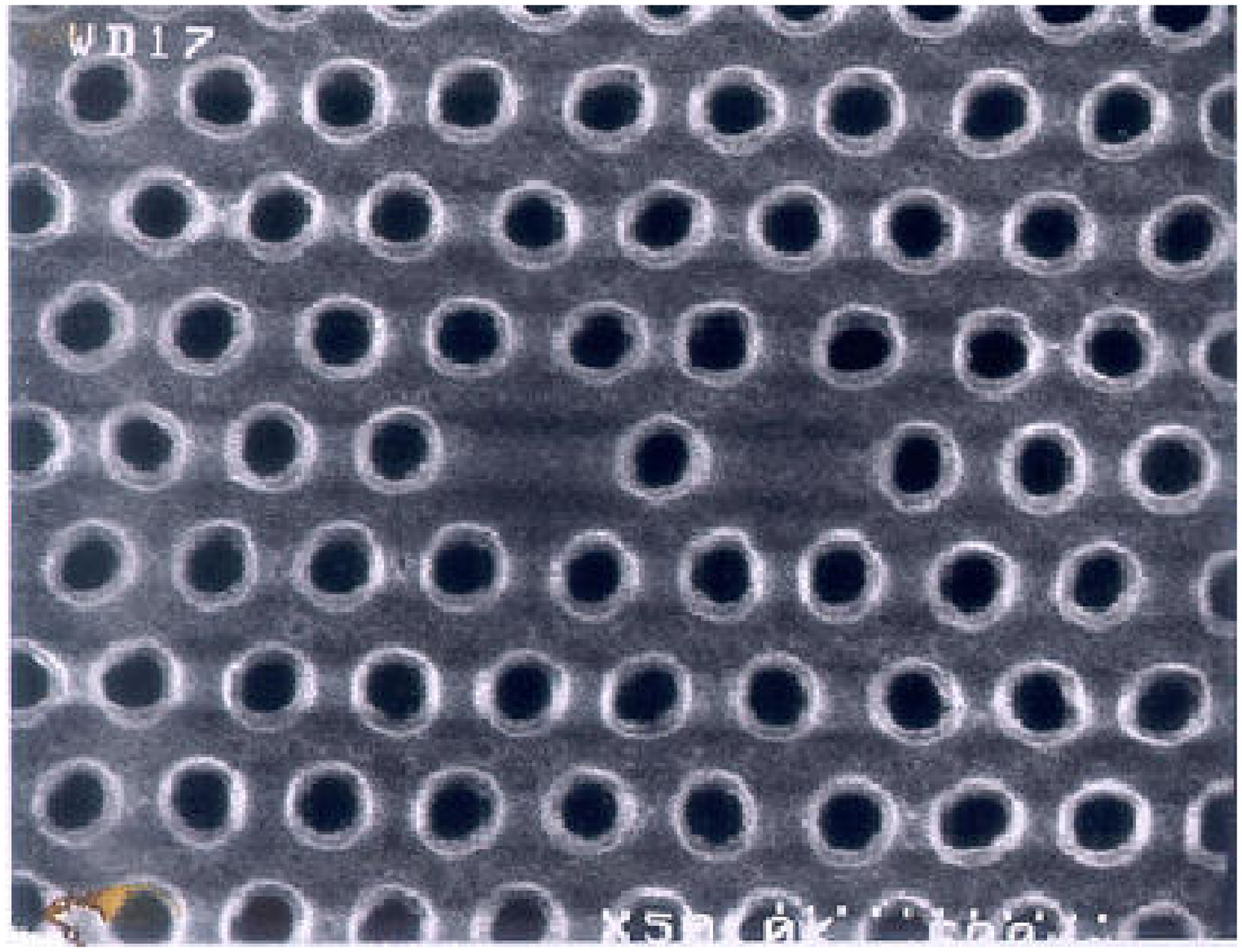, width=2.4in}}
\caption{SEM pictures showing the top
views of the fabricated structures.} \label{fig:procwaf2}
\end{center}
\end{figure}

The mode pattern of the constructively coupled defect state is shown in Figure \ref{fig:spregydips}. Parameters of the mode are: $a/\lambda=0.288$, $Q_{\perp}=12120$,
$V_{mode}=0.14(\lambda/2)^3$, $m_0=1.4\cdot 10^{-7}$ and $N_0=0.0063$. Strong coupling is achievable for an atom trapped in any of the two central holes of the
unperturbed radius (positioned between the defects). For $\lambda=852nm$, this radius is again $r=68nm$.

\section{Fabrication}

We have recently developed the fabrication procedure for making
these cavities in $AlGaAs$. The material and PC properties are
chosen in such a way that cavities operate at $\lambda=852nm$ (the
wavelength corresponding to the atomic D2 transition in
$^{133}Cs$).

The fabrication process starts by spinning of $100nm$ thick high molecular weight PMMA (polymethylmethacrylate) on top of the wafer. The PMMA layer is
subsequently baked on a hot plate at $150^o C$ for 20 minutes. A desired 2D PC pattern is beamwritten on the PMMA by electron beam lithography in a Hitachi
S-4500 electron microscope. The exposed PMMA is developed in a 3:7 solution of 2-ethoxyethanol:methanol for 30 seconds. The pattern is then transferred into the
$AlGaAs$ layer using the $Cl_2$ assisted ion beam etching. After that, the sacrificial $AlAs$ layer is dissolved in hydrofluoric acid (HF) diluted in water. HF
attacks $AlAs$ very selectively over $Al_xGa_{1-x}As$ for $x<0.4$ \cite{ref:Yablo90}. Therefore, the percentage of $Al$ in our $AlGaAs$ layer is around $30\%$.
Finally, the remaining PMMA may be dissolved in acetone.

Three SEM pictures showing top views of fabricated microcavity
structures are shown in
Figure \ref{fig:procwaf2}. We are currently working to measure the
passive optical properties of such microcavities in order to
validate our theoretical predictions.

\section{Conclusion}

In conclusion, we have theoretically demonstrated that PC cavities
can be designed for strong interaction with atoms trapped in one
of PC holes. At present, we are working on further optimization of
the design and the characterization of fabricated structures.

Critical issues for further investigation include efficient
coupling of light in and out of the PC microcavity, as well as
accurate estimation of surface effects that could perturb atomic
radiative structure within the small defect hole. The extremely
small mode volume in these structures also poses an interesting
theoretical question of how standard cavity QED models must be
modified when the single-photon Rabi frequency exceeds the atomic
hyperfine spacing.

\section{Appendix: The effect of fractional edge dislocations}

Any wavefront can be considered as a source of secondary waves
that add to produce distant wavefronts, according to Huygens
principle. Let us assume that we know the field distribution
accross the plane $S$, positioned in the near field, above the
free standing membrane and in parallel to the membrane surface.
The far fields can be considered as arising from the equivalent
current sheets on this plane. Therefore, we can calculate the far
field distribution and the total averaged radiated power into the
half-space above the plane $S$\cite{ref:RWVD}:

\begin{equation}
\begin{split}
P&=\frac{\eta}{8\lambda^2}\int\limits_0^{\pi/2} \int\limits_{0}^{2\pi} d_\theta d_\phi sin(\theta) K(\theta,\phi)\\
K(\theta,\phi)&=\biggl | N_\theta+\frac{L_\phi}{\eta}\biggr |^2 +
\biggl | N_\phi-\frac{L_\theta}{\eta}\biggr |^2 \\
\eta&=\sqrt{\frac{\mu_o}{\epsilon_o}},
\end{split}
\label{equation:P1}
\end{equation}

where $\vec{N}$ and $\vec{L}$ represent radiation vectors, whose
components in the rectangular coordinate system are proportional
to Fourier transforms of tangential field components at the plane
$S$\cite{ref:JV2001new}:

\begin{align}
N_x&=-FT_2(H_y)\biggr |_{\vec{k}_{||}}\\
N_y&=FT_2(H_x)\biggr |_{\vec{k}_{||}}\\
L_x&=FT_2(E_y)\biggr |_{\vec{k}_{||}}\\
L_y&=-FT_2(E_x)\biggr |_{\vec{k}_{||}}\\
\vec{k}_{||}&=\frac{2\pi}{\lambda}sin\theta(\hat{x}cos\phi+\hat{y}sin\phi)\\
FT_2(f(x,y))&=\iint \limits_S d_x d_y f(x,y) e^{i (k_x x + k_y
y)},
\end{align}

\begin{figure}[hbtp]
\begin{center}
\epsfig{figure=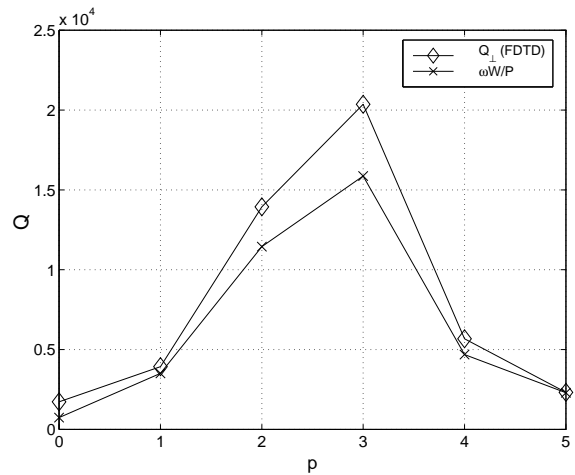, width=3in}
\caption{Q factor computed
using the FDTD method, and from the expression \ref{equation:P1},
for the structure analyzed in the subsection
\ref{subsection:n24addp}.} \label{fig:rf}
\end{center}
\end{figure}

Therefore, just by knowing the Fourier transforms of the tangential field components at the plane $S$, we can evaluate the time averaged radiated power $P$. From the previous expressions it is clear that the wave-vector of interest $\vec{k}_{||}$ lies within the light cone for any values of angles $\theta$ and $\phi$ in the circular polar coordinate system (i.e. $|\vec{k}_{||}|\leq \frac{2\pi}{\lambda}$). This implies that the radiated power $P$ depends only on the wave-vector components located within the light cone. It is also clear that by suppressing the Fourier components within the light cone, one can reduce $P$. Having in mind that for the $x$-dipole mode $E_y$ and $H_x$ fields are odd with respect to both $x$ and $y$ symmetry axes, $N_y$ and $L_x$ do not contribute significantly to the integral in expression \ref{equation:P1} (they are both equal to zero at any point in the k-space with either $k_x$ or $k_y$ equal to zero). On the other hand, $E_x$ and $H_y$ field components are even with respect to both $x$ and $y$ axes and their Fourier transforms are generally non-zero at small wave-vector values. However, by tuning the elongation factor $p$, one can balance the energy in the positive and negative field lobes and minimize the Fourier components of $E_x$ and $B_y$ within the light cone. This also leads to a decrease in the radiated power $P$. We can conclude that the improvement in the $Q$ factor after the application of fractional edge dislocations is due to the suppression of the wave-vector components composing the defect mode, which are located within the light cone. A more detailed explanation of this phenomenon and how it can be used to improve $Q$ factors of other types of modes will be presented in our forthcoming publication\cite{ref:JV2001new}. \\
\indent The $Q$ factor of a mode can be expressed as
$Q=\omega\frac{W}{P}$, where $W$ is the total energy of a mode in
the half-space $z\geq 0$. The comparison between the $Q$ factor
calculated using the method presented in this appendix, and
$Q_\perp$ previously estimated using the FDTD, for the structure
from the subsection \ref{subsection:n24addp}, is shown in Figure
\ref{fig:rf}. The plane $S$ is positioned directly above the
surface of the membrane in this case. From Figure
\ref{fig:n24addp3} it follows that the total $Q$ factor saturates
at about $17000$, when the number of PC layers around the defect
increases in the structure with $p=3$. This is very close to the
maximum $Q$ value estimated from the expression \ref{equation:P1}.

\begin{bf}Acknowledgement \end{bf}\\\\
This work has been supported by the Caltech MURI Center for Quantum Networks.

\bibliographystyle{unsrt}
\bibliography{qc_ref}

\begin{thebibliography}{10}

\bibitem{ref:Painter99}
O.~Painter, R.K. Lee, A.~Scherer, A.~Yariv, J.D. O'Brien, P.D. Dapkus, and
  I.~Kim.
\newblock {Two-Dimensional Photonic Band-Gap Defect Mode Laser}.
\newblock {\em Science}, 284:1819--1821, June 1999.

\bibitem{ref:Loncar00}
M.~Lon\v{c}ar, D.~Nedeljkovi\'{c}, T.~Doll, J.~Vu\v{c}kovi\'{c}, A.~Scherer,
  and T.~P. Pearsall.
\newblock {Waveguiding at 1500nm using photonic crystal structures in silicon
  on insulator wafers}.
\newblock {\em Applied Physics Letters}, 77:1937--1939, September 2000.

\bibitem{ref:Kimb94}
H.~J. Kimble.
\newblock {\em {in {\it Cavity Quantum Electrodynamics}, edited by P. Berman}}.
\newblock Academic Press, San Diego, 1994.

\bibitem{ref:Painter98}
O.~Painter, J.~Vu\v{c}kovi\'{c}, and A.~Scherer.
\newblock {Defect Modes of a Two-Dimensional Photonic Crystal in an Optically
  Thin Dielectric Slab}.
\newblock {\em Journal of the Optical Society of America B}, 16(2):275--285,
  February 1999.

\bibitem{ref:Yablo91}
E.~Yablonovitch, T.J. Gmitter, R.D. Meade, A.M. Rappe, K.D. Brommer, and J.D.
  Joannopoulos.
\newblock {Donor and acceptor modes in photonic band-structure}.
\newblock {\em Physical Review Letters}, 67:3380--3383, 1991.

\bibitem{ref:JV_spie2001}
J.~Vu\v{c}kovi\'{c}, M.~Lon\v{c}ar, and A.~Scherer.
\newblock {Design of photonic crystal optical microcavities}.
\newblock {\em Proceedings of SPIE, Photonix West 2001}, 2001.

\bibitem{ref:Weinstein95}
J.~D. Weinstein and K.~G. Libbrecht.
\newblock {}.
\newblock {\em Physical Review A}, 52:4004, 1995.

\bibitem{ref:Drndic01}
M.~Drndi\'{c}, C.~S. Lee, and R.~M. Westervelt.
\newblock {Three-dimensional microelectromagnet traps for neutral and charged
  particles}.
\newblock {\em Physical Review B}, 63:085321, 2001.

\bibitem{ref:Haensch}
J.~Reichel, W.~Hansel, P.~Hommelhoff, and T.~W. Hansch.
\newblock {Applications of integrated magnetic microtraps}.
\newblock {\em Applied Physics B}, 72:81, 2001.

\bibitem{ref:Juerg}
M.~Bartenstein, D.~Cassettari, T.~Calarco, A.~Chenet, R.~Folman, K.~Brugger,
  A.~Haase, E.~Hartungen, B.~Hessmo, A.~Kasper, P.~Kruger, T.~Maier, F.~Payr,
  S.~Schneider, and J.~Schmiedmayer.
\newblock {Atoms and wires: Toward atom chips}.
\newblock {\em IEEE Journal of Quantum Electronics}, 36:1364, 2000.

\bibitem{ref:Barton97}
G.~Barton.
\newblock {Van der Waals shifts in an atom near absorptive dielectric mirrors}.
\newblock {\em Proceedings of the Royal Society of London A}, 453:2461--2495,
  1997.

\bibitem{ref:Chevrollier92}
Martine Chevroliier, Michele Fichet, Marcos Oria, Gabriel Rahmat, Daniel Bloch,
  and Martial Ducloy.
\newblock {High resolution selective reflection spectroscopy as a probe of
  long-range surface interaction: measurement of the surface van der Waals
  attraction exerted on excited Cs atoms}.
\newblock {\em Journal de Physique II}, 2:631--657, 1992.

\bibitem{ref:Yablo90}
E.~Yablonovitch, D.M. Hwang, T.J. Gmitter, L.T. Florez, and J.P Harbison.
\newblock {Van der Waals Bonding of GaAs Epitaxial Liftoff Films onto Arbitrary
  Substrates}.
\newblock {\em Applied Physics Letters}, 56(24):2419--2421, June 1990.

\bibitem{ref:RWVD}
S.~Ramo, J.R. Whinnery, and T.~Van Duzer.
\newblock {\em {Fields and waves in communication electronics}}.
\newblock John Wiley and Sons, Inc., New York, 1994.

\bibitem{ref:JV2001new}
J.~Vu\v{c}kovi\'{c}, M.~Lon\v{c}ar, H.~Mabuchi, and A.~Scherer.
\newblock {Optimization of Q-factor in microcavities based on free-standing
  membranes}.
\newblock {\em IEEE Journal of Quantum Electronics}, 38(7):850--856, July 2002.

\end{thebibliography}

\end{document}